\documentclass[aip,amsmath,amssymb,reprint]{revtex4-1}

\usepackage{graphicx}
\usepackage{bm}

\usepackage{color}

\begin{document}

\title{Size-dependence of hydrophobic hydration at electrified gold/water interfaces}
\author{Alessandra Serva}
\affiliation{Sorbonne Universit\'e, CNRS, Physico-chimie des Electrolytes et Nanosyst\`emes Interfaciaux, PHENIX, F-75005 Paris, France}
\author{Mathieu Salanne}
\affiliation{Sorbonne Universit\'e, CNRS, Physico-chimie des Electrolytes et Nanosyst\`emes Interfaciaux, PHENIX, F-75005 Paris, France}
\affiliation{Institut Universitaire de France (IUF), 75231 Paris Cedex 05, France}
\author{Martina Havenith}
\affiliation{Department of Physical Chemistry II, Ruhr University Bochum, 44780 Bochum, Germany}
\author{Simone Pezzotti}
\email{simone.pezzotti@rub.de}
\affiliation{Department of Physical Chemistry II, Ruhr University Bochum, 44780 Bochum, Germany}


\begin{abstract}

Hydrophobic hydration at metal/water interfaces actively contributes to the energetics of electrochemical reactions, e.g. CO$_2$ and N$_2$ reduction, where small hydrophobic molecules are involved. In this work, constant applied potential molecular dynamics is employed to study hydrophobic hydration at a gold/water interface. We propose an extension of the Lum-Chandler-Weeks (LCW) theory to describe the free energy of hydrophobic hydration at the interface as a function of solute size and applied voltage. Based on this model we are able to predict the free energy cost of cavity formation at the interface directly from the free energy cost in the bulk plus an interface-dependent correction term. The interfacial water network contributes significantly to the free energy yielding a preference for outer-sphere adsorption at the gold surface for ideal hydrophobes. We predict an accumulation of small hydrophobic solutes of sizes comparable to CO or N$_2$, while the free energy cost to hydrate larger hydrophobes, above 2.5 \AA~radius, is shown to be greater at the interface than in the bulk. Interestingly, the transition from the volume dominated to the surface dominated regimes predicted by the LCW theory in the bulk is also found to take place for hydrophobes at the Au/water interface, but occurs at smaller cavity radii. By applying the extended LCW theory to a simple model addition reaction, we illustrate some implications of our ﬁndings for electrochemical reactions.

\end{abstract}

\maketitle

\section{Introduction}

Small hydrophobic species are often present at metal/water interfaces, as reactants, intermediates and products in a large variety of electrochemical processes, such as CO$_2$\cite{Norskov_CHEMCATCHEM2015,Zhao_ACSENLET2018} and N$_2$ reduction~\cite{Bao_ADVMAT2017,skulason_PCCP2012}.
The development of models able to describe hydrophobic hydration at the interface with a metal is therefore a key step in the optimization of these reactions.\\
\indent The structure and the dynamics of water molecules adsorbed at metallic surfaces is now well understood thanks to the combination of advanced experimental (e.g. surface specific in-situ vibrational spectroscopies and synchrotron-based techniques)~\cite{Shen_PNAS2014,Velasco_SCIENCE2014,Tong_ANGEW2017,Cheng_NATMAT2019} and computational~\cite{Michaelidis_PRL2003,Willard_FARADISC2009,Steinman_JCTC2020,Cheng_NATMAT2019,Cheng_SciAdv2020,Gros_JCP2018,ROSSMEISL_CPL2208,Hansen_CHEMSCI2018,Cheng_JACS2018} methods. A very interesting result arising from these studies is the existence of hydrophobic effects at the interface due to the peculiar organization of the hydrogen bond (HB) network of the adsorbed water molecules~\cite{Limmer_PNAS2013}.  Hydrophobic hydration is a key phenomenon in bulk water~\cite{Chandler_JPCB1999,Chandler_NAT2005}, which understanding has lead to important progresses in e.g. our comprehension of biological processes where it is ubiquitous~\cite{Chandler_NAT2005,Garde_fluctuationsREVIEW2011,Richard_JPCB2018,Patel_JPCB2014,Patel_PNAS2011,Rossky_PNAS2008,Rossky_NATURE1998}.
A molecular description of hydrophobicity is still a challenge for theory and experiments~\cite{Hassanali_NatCom2020,Garde_fluctuationsREVIEW2011,Chandler_PNAS2000,Zangi_JACS2009,Pezzotti_JPCL2020}.\\
\indent As described by the Lum-Chandler-Weeks (LCW) theory~\cite{Chandler_JPCB1999}, the free energy cost of solvating a hydrophobic solute in bulk water is well approximated by the free energy cost to form a cavity in the liquid, and primarily depends on the cavity size. In particular, for small solutes ($<$7~\AA~radius), the water HB network responds elastically to accommodate the cavity, but a reduced number of intact network configurations results in an entropy decrease proportional to the volume. On the other hand, interfacial thermodynamics applies for larger solutes, involving the breaking of water-water HBs and a corresponding change in enthalpy that scales with the cavity surface area~\cite{Chandler_NAT2005}. When moving from the bulk to the interface, the hydration of hydrophobic species is further modulated by specific water-water and water-surface interactions, and deviations from the bulk behavior are often observed~\cite{Garde_fluctuationsREVIEW2011,Limmer_PNAS2013, Garde_PNAS2009, Patel_PNAS2011}. For instance, large density fluctuations of the water surface in contact with hydrophobic media were shown to promote the solvation of hydrophobic solutes, while a more bulk-like behaviour has been reported for hydrophilic interfaces, where density fluctuations are suppressed by water-surface interactions~\cite{Chandler_JPCB2010,Garde_fluctuationsREVIEW2011}. \\
\indent In this respect, metal/water interfaces represent a special case, where strong water-surface interactions lead to very ordered water adlayers on top of the metal, with a soft liquid interface being however formed between the adlayer and the adjacent water layer~\cite{Limmer_PNAS2013,Willard_FARADISC2009,Michaelidis_ChemRev2016}. In the case of platinum, this water-water interface was shown to exhibit density fluctuations typical of hydrophobic environments, promoting the formation of cavities that can accommodate small solutes~\cite{Limmer_PNAS2013}. In the present work, we investigate hydrophobic hydration at the electrified interface between water and a gold (100) surface by means of classical molecular dynamics (MD) simulations. By focusing on a range of solute size and surface potential relevant for electrochemical applications, we provide a surface-dependent correction to the LCW theory that accounts for the modifications imposed by the gold surface on the cavity formation mechanism, and we shed light on its dependence on the applied voltage.

\begin{figure*}[]
\begin{center}
\includegraphics[width=0.8\textwidth]{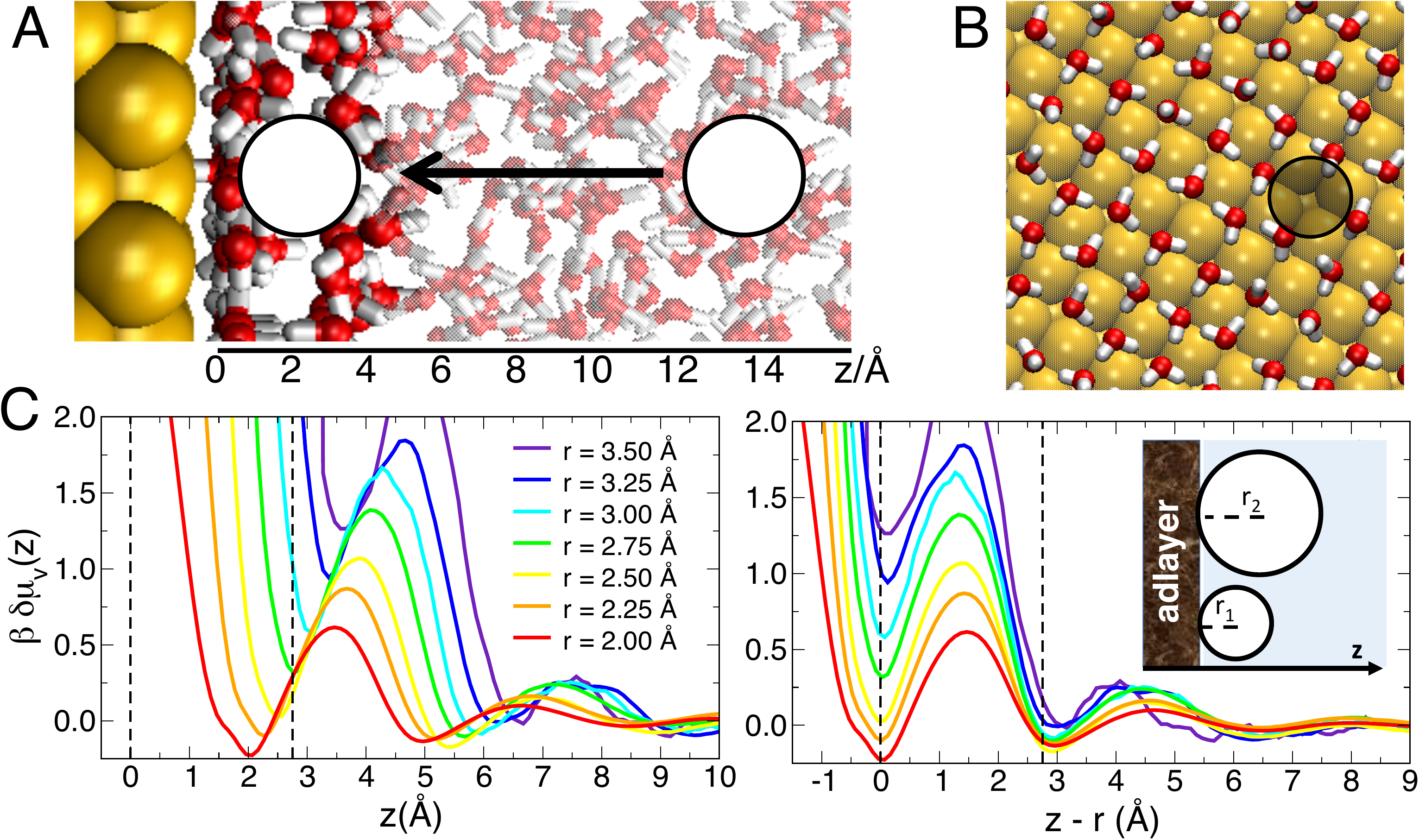}\\
\end{center}\caption{\textbf{Excess solvation free energy for ideal hydrophobes at the Au(100)/water interface at PZC}. A) Schematic illustrating how the excess solvation free energy, $\delta\mu_v(z)$, with v the volume of the ideal hydrophobe (white sphere) and z the vertical distance from the adlayer, is calculated from MD simulations. The water molecules belonging to the adlayer (in direct contact with Au, at z$\sim$0 \AA) and to the adjacent air/water-like layer (at z$\sim$3 \AA) are highlighted. The white spheres identify the hydrophobe solvated in the bulk and in its most stable position at the interface, i.e. with one side contacting the adlayer. B) Top view of the ordered water arrangement within the adlayer, showing a square symmetry with some vacancies (see black circle for one example). C) Left: $\delta\mu_v(z)$ vs $z$ profiles for ideal hydrophobes with increasing radius (r). Right: Same profiles plotted as a function of $z-r$. The vertical dashed lines identify the location of the adlayer and the air/water-like layer. }\label{deltamu}
\end{figure*}
\section{Results}\label{Result}
\subsection{Hydrophobic hydration at the interface} 
Fig. \ref{deltamu} reports the solvation free energy profiles $\delta\mu_v(z)$, where $z$ is the vertical distance from the Au surface, for small spherical hydrophobic solutes with increasing radius (from $r$~=~2.0~\AA~to $r$~=~3.5~\AA). The profiles are obtained by monitoring the probability to form spherical cavities of
the chosen radius as a function of $z$ (Eq. \ref{eqn:free_energy}, methods section). The applied voltage ($\Delta V$) is set to 0~V, which corresponds to the point of zero charge (PZC) of the model. A characteristic snapshot illustrating this methodology is provided in Fig. \ref{deltamu}A, where the interfacial organization of water molecules is also highlighted. This latter resembles the one previously shown by Limmer et al. for Pt/water interfaces~\cite{Limmer_PNAS2013}: strong, favorable interactions between the water molecules and the metal surface lead to peculiar interactions with adjacent water molecules. As illustrated in the top view of panel B, water molecules within the 1$^{st}$ adsorbed layer (the adlayer) preferentially lie flat on the surface in a distance of $\sim$3 \AA. The global arrangement exhibits a square symmetry, with however some vacancies left, which position is dynamic in time. Similar vacancies were observed in the monolayer structures of water absorbed on many face-centered cubic (FCC) metals~\cite{Salmeron_JACS2009}. The ordered water structure above the Au surface leads to a maximization of the number of HBs formed between adlayer water molecules (19 HBs/nm$^2$ on average; see Methods for HBs definition). As a consequence, few spots remain available for forming HBs between the adlayer and the 2$^{nd}$ water layer, resulting in only 4 inter-layer HBs/nm$^2$. For this reason, the 2$^{nd}$ layer was shown to resemble the one formed by water in contact with hydrophobic media~\cite{Limmer_PNAS2013}, e.g. air, and we refer to it as the air/water-like layer. The hydrophobic character of the interface between the adlayer and the air/water-like layer is responsible for large water density fluctuations, increasing the probability to form cavities. This appears clearly on the hydration free energy profiles in Fig.~\ref{deltamu}C-left, which show a minimum at $z$~$<$~5~\AA~for all the investigated cavity radii. The minima represent stable locations for hydrophobic solutes at the interface. \\
\begin{figure*}[]
\begin{center}
\includegraphics[width=0.9\textwidth]{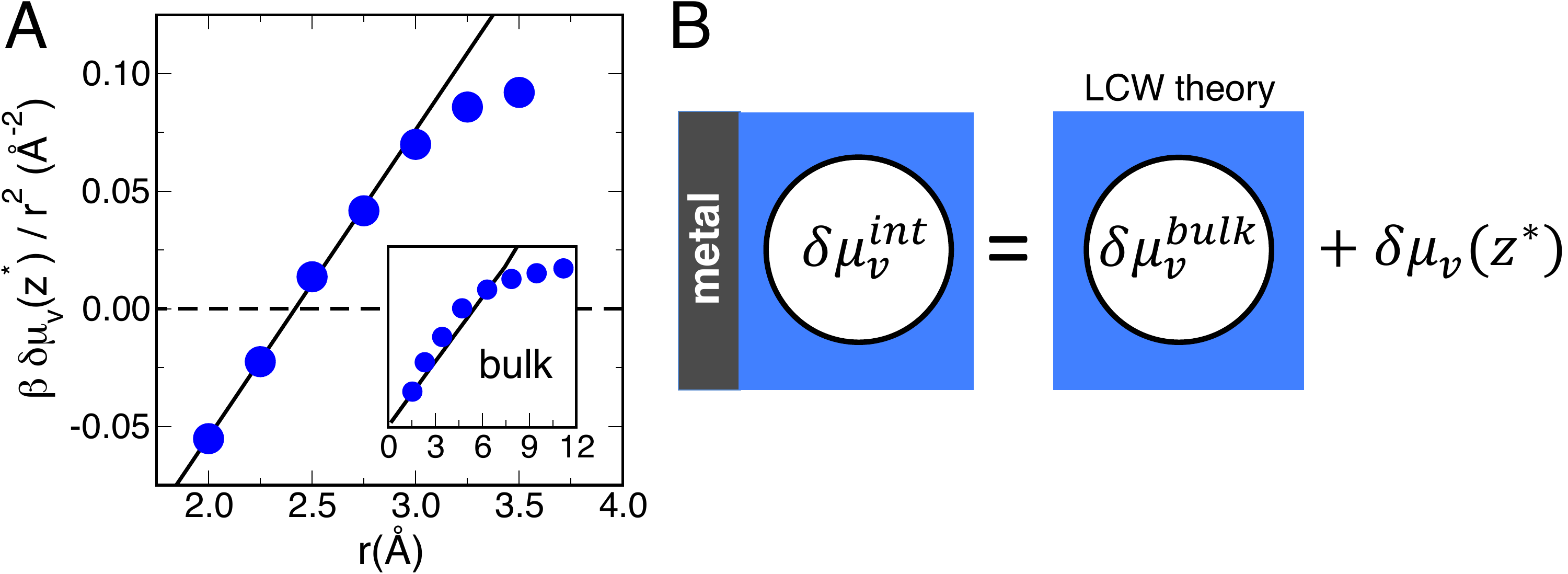}\\
\end{center}\caption{\textbf{Hydrophobic hydration at the Au(100)/water interface versus bulk.} A) Excess solvation free energy for ideal hydrophobes as a function of their radius, $r$, normalized by the cavity surface area. The reported $\delta\mu_v$ values are differences between the solvation free energy of the hydrophobe located at the interface (z=z*, first minimum in Fig. \ref{deltamu}C) and in the bulk. The black solid line is a linear fit, highlighting the volume-dominated regime strictly valid until r=3.0 \AA. The inset reports the plot obtained from the LCW theory~\cite{Chandler_JPCB1999} for bulk water, adapted from Ref.\cite{Chandler_NAT2005}. B) $\delta\mu_v(z^*)$ provides the means to obtain the free energy cost of cavity formation at the interface, $\delta\mu_v^{int}$, directly from the well-known values in the bulk, $\delta\mu_v^{bulk}$.  }\label{chandler}
\end{figure*}
\indent When the cavity radius increases, the free energy minimum is located further away from the surface and its value increases. To rationalize this result, we should consider not only the distance between the center of the cavity and the Au surface, but rather the minimum distance between the cavity surface and Au, $z-r$, as reported in the right panel of Fig. \ref{deltamu}C. For all cavities, the most stable location at the interface is at $z-r=$~0~\AA, i.e. when the cavity surface contacts the adlayer in one point, as illustrated in the inset. This corresponds to an outer-sphere adsorption of the hydrophobes, which are separated from the metal by the adlayer. By contrast, inner-sphere adsorption, that would require the formation of a cavity within the adlayer, is largely disfavoured, as measured by the sharp increase of $\delta\mu_v(z)$ for $z-r$~$<$~0~\AA. This can be understood by considering that replacing a water molecule in the adlayer with an empty patch has a high free energy cost, which has been estimated in Ref.\cite{Willard_FARADISC2009} to be on the order of 10~k$_B$T. More details regarding the features in the $\delta\mu_v(z)$ profiles can be found in the Supporting Information (SI, Fig. S1 and related discussion).\\

\subsection{Extended LCW theory for the metal-water interface}
The value of $\delta\mu_v(z)$ at the minimum ($z-r =$~0~\AA) is denoted hereafter $\delta\mu_v(z^*)$.
This term corresponds to the difference in the free energy cost to form the same cavity in the most stable position at the interface and in the bulk: 
\begin{equation}
\delta\mu_v(z^*) = \delta\mu_v^{int} - \delta\mu_v^{bulk} \label{eqn:LCW}
\end{equation}
Therefore, it provides a correction term that allows to extend the LCW theory, which describes the hydration free energy of ideal hydrophobes in the bulk~\cite{Chandler_NAT2005,Chandler_JPCB1999}, to the Au/water interface. Fig. \ref{chandler}A shows the variations of $\delta\mu_v(z^*)$  normalized by the cavity surface area  with respect to the cavity radius. This representation allows to establish a proportional increase of the free energy to the cavity volume, as it is the case in the bulk~\cite{Chandler_NAT2005,Chandler_JPCB1999} for cavity radii smaller than 7~\AA\ (see the inset of the Figure). A similar volume-dominated regime is observed for the additional interfacial contribution, as shown by the linear dependence of $\delta\mu_v(z^*)/r^2$ on $r$. However, the linear behavior is strictly valid for a more limited radii range, i.e. until $r$~=~3.0~\AA, in contrast to the bulk case.\\
\indent Moreover, as observed in the plot, $\delta\mu_v(z^*)/r^2$ is equal to zero at about $r$~=~2.5~\AA, which according to Eq.\ref{eqn:LCW} corresponds to the case when the cavity formation at the interface and in the bulk are isoenergetic, while it becomes less than 0 for $r$~$<$~2.5~\AA. Therefore, the formation of small cavities, that can accommodate the smallest hydrophobic molecules such as N$_2$ or CO, is favoured at the interface with respect to the bulk, promoting their accumulation near the Au surface. By contrast, $\delta\mu_v(z^*)$ becomes positive for cavities with $r$~$>$~2.5~\AA, representative of larger hydrophobic molecules such as CO$_2$ or CH$_4$. Thus, a substantial accumulation of these molecules at the interface is not promoted by the water network, even though the well-defined minimum at $z-r=$~0~\AA~in Fig.~\ref{deltamu}C demonstrates that they are metastable at the metal-water interface. Interestingly, the crossover from negative to positive $\delta\mu_v(z^*)/r^2$ values goes beyond what is expected for canonical hydrophobic interfaces, where density fluctuations are enhanced with respect to bulk in both small and large observation volumes~\cite{Chandler_JPCB2010,Garde_fluctuationsREVIEW2011}. The water-water interface formed between the adlayer and the air/water-like layer hence plays a dual role: on the one hand it creates a hydrophobic environment where short-range density fluctuations that can accommodate small hydrophobic solutes are enhanced; on the other hand it constrains long-range density fluctuations, so that the formation of large cavities is hindered.\\
\begin{figure}[]
\begin{center}
\includegraphics[width=0.47\textwidth]{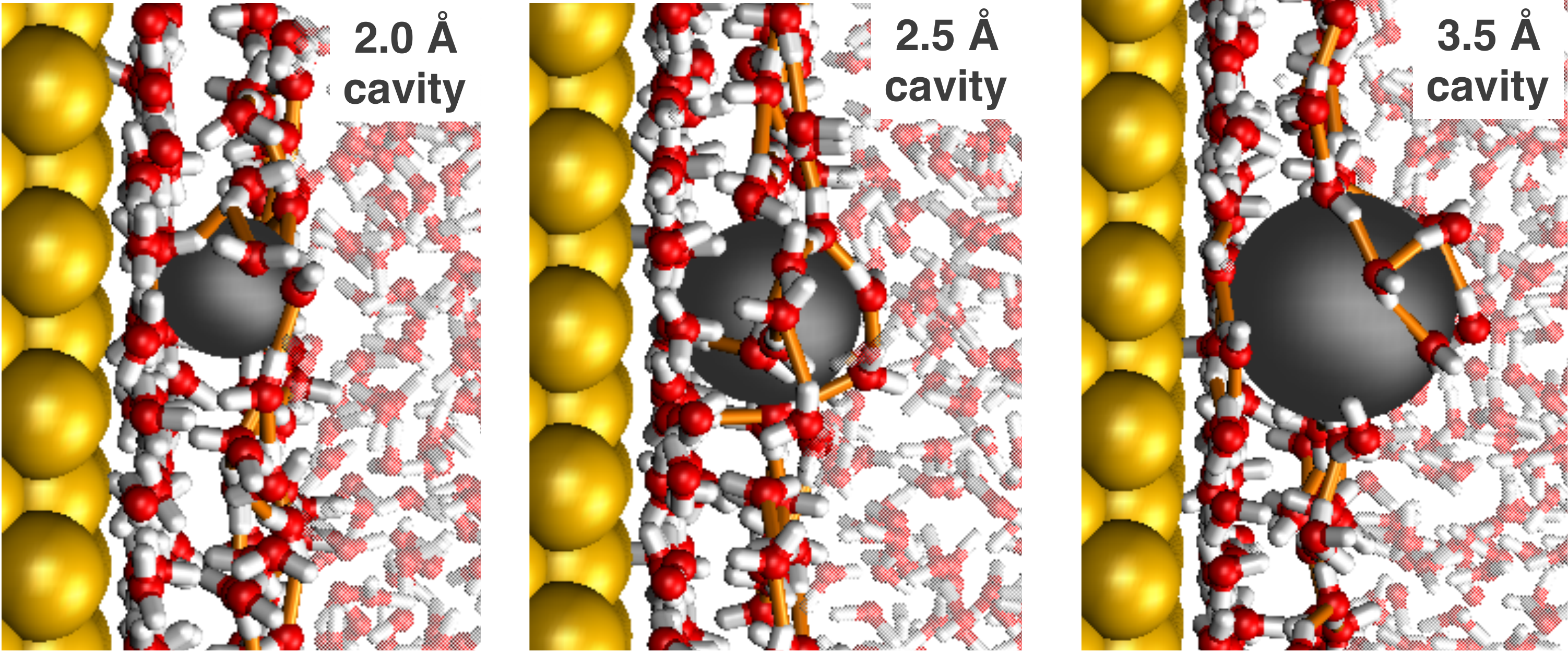}\\
\end{center}\caption{\textbf{Hydrophobic hydration mechanism at the Au(100)/water interface at PZC.} MD snapshots illustrating the mechanism for the solvation of small ideal hydrophobes (i.e. small cavities, grey spheres) in their most stable position at the interface (first minimum in Fig. \ref{deltamu}C). The HBs between water molecules in the air/water-like layer are materialized in orange to better appreciate the distorsion (for 2.5 \AA~in radius cavities) and the local breaking (for 3.5 \AA~in radius cavities) of the interfacial water network.}\label{cavity_snap}
\end{figure}
\indent In order to better elucidate the microscopic origin of such dual role, a direct visualization of the cavities formed during the simulation is provided in Fig. \ref{cavity_snap}. For the smallest cavities, represented in the Figure by the $r=2.0$~\AA\ panel, most of the empty volume occupies the inter-layer space in between the adlayer and the air/water-like layer (which are separated by $\sim$3 \AA), and the HB-network within the air/water-like layer (orange bonds in the figure) is virtually unperturbed around the cavity. Since the density of HBs between the adlayer and the air/water-like layer is extremely low (only 4 HBs/nm$^2$), the free energy cost of cavity formation is minimized in the inter-layer space, thus $\delta\mu_v(z^*)/r^2<0$. However, when the size of the cavity increases above 2.5~\AA, i.e. when we enter in the size-range where $\delta\mu_v(z^*)/r^2$ $>0$ and still scales linearly with the cavity volume, the distortion imposed on the HB-network in the air/water-like layer becomes more severe (see orange bonds in the middle panel). As it can be seen in the snapshot, this distortion arises from an elastic response of the water network, which is wrapped around the cavity. According to the LCW theory~\cite{Chandler_JPCB1999}, this process has an entropic cost, that is found to be larger at the interface than in bulk water since $\delta\mu_v(z^*)/r^2>0$. Thus, the ``flexibility'' of the hydrophobic air/water-like layer is reduced in presence of the adlayer, as this latter can provide some HBs, even if few, that locally pin water density fluctuations. This is not the case for purely hydrophobic interfaces, which accordingly promote the accommodation of both small and large hydrophobic solutes~\cite{Chandler_JPCB2010,Garde_fluctuationsREVIEW2011}. The effect of the adlayer becomes more pronounced when increasing the cavity radius up to 3.5~\AA~(Fig. \ref{cavity_snap}, right panel), where the HB-network within the air/water-like layer is locally broken (missing orange bonds in the Figure) in proximity of the cavity. The breaking of some HBs introduces an enthalpic component to the free energy cost of cavity formation, which according to the LCW theory is expected to scale with the cavity surface area. This is the reason why deviations from the volume-dominated regime in Fig. \ref{chandler} start to be observed for cavities of 3.25 and 3.5~\AA~radius. Our results therefore demonstrate that the transition from the volume dominated (entropic) to the surface dominated (enthalpic) regimes predicted by the LCW theory in the bulk also takes place at the Au/water interface, but it is anticipated to smaller cavity radii due to the constraints imposed by the adlayer on the fluctuations of the interfacial water network.\\

\subsection{Effect of applied potential}
\begin{figure}[t!]
\begin{center}
\includegraphics[width=0.47\textwidth]{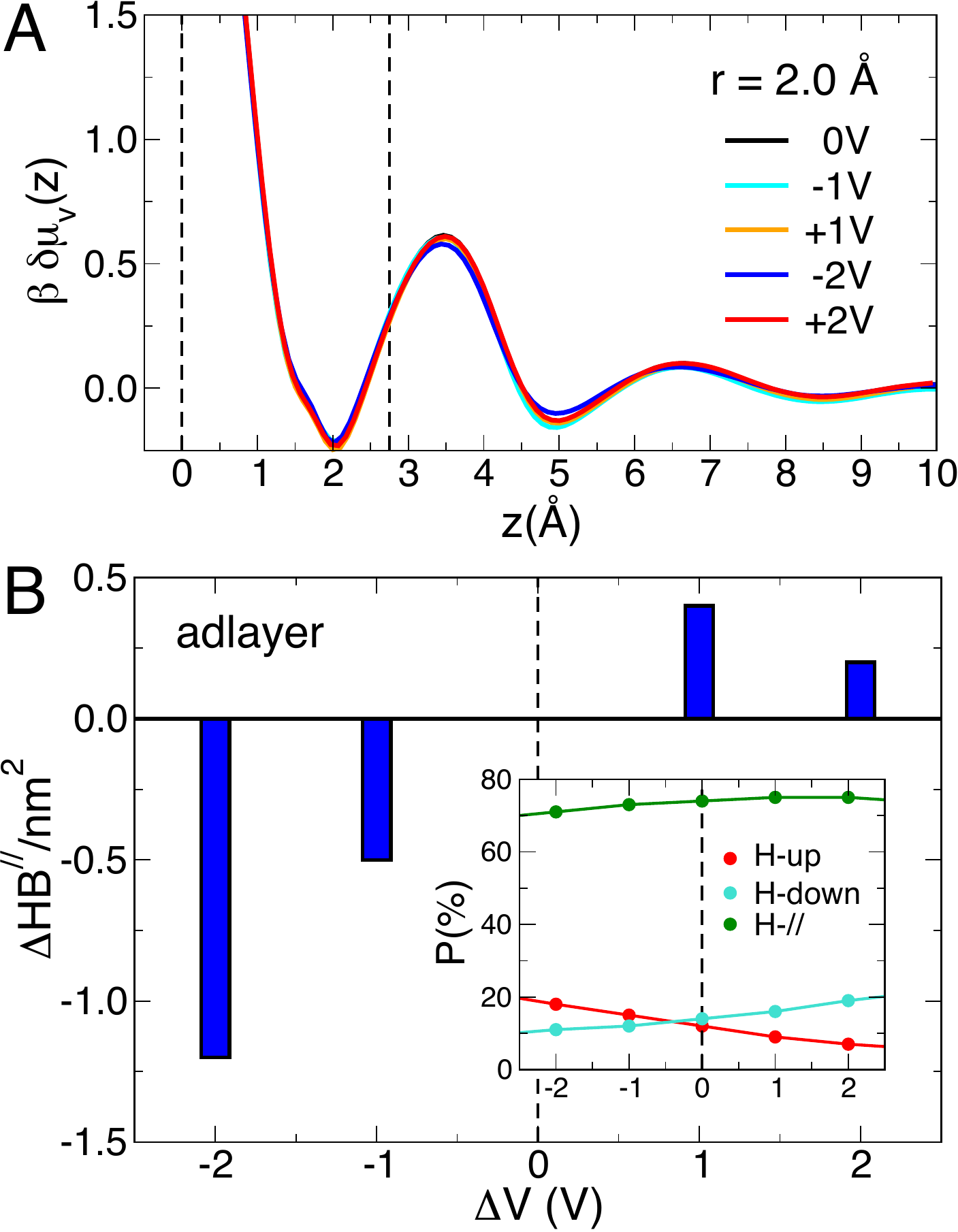}\\
\end{center}\caption{\textbf{Effect of the applied potential.} A) Excess solvation free energy, $\delta\mu_v(z)$, for a 2.0 \AA~radius ideal hydrophobe at negative and positive electrodes. Similar trends for a larger cavity are shown in the SI, Fig. S2. B) Adlayer water structure as a function of the applied voltage, as quantified by the variation in the number of HBs ($\Delta HB^{//}$) formed between adlayer water molecules with respect to PZC. The inset shows the probability for adlayer waters to orient their hydrogen atoms toward the Au surface (H-up, red), away from the Au surface (H-down, cyan) or parallel to it (H-//, green). Water density profiles are shown in the SI,  Fig. S3.}\label{volt1}
\end{figure}

In Fig.~\ref{volt1}, we now evaluate the effect of an applied voltage by varying the Au slab potential from -2~V to +2~V. Since interfacial chemical reactions are not included in our classical simulations, only the potential-induced structural changes within the adlayer and the air/water-like layer will affect density fluctuations and cavity formation at the interface. Fig.~\ref{volt1}A demonstrates that the effect of the surface potential on the free energy cost of cavity formation is negligible in the investigated range, with the free energy profiles obtained at all potential values being almost superimposed. Therefore, all the previous results obtained for PZC conditions can be extended to a wide range of working electrochemical conditions, as long as no reactions occur.\\
\indent Despite hydrophobic hydration remaining unaltered, structural changes within the adlayer are detected upon voltage application. In Fig.~\ref{volt1}B they are evaluated by quantifying the variation in the number of HBs formed between adlayer water molecules, denoted HBs$^{//}$ due to their orientation parallel to the Au surface, as well as by the changes in the orientation of adlayer water OH-groups (inset). At 0~V (PZC), adlayer water molecules preferentially lie parallel to the Au surface ($>$70\% of the OH-groups with parallel orientation, green curve in the inset), and form 19 HBs$^{//}$/nm$^2$. Once a negative potential is applied, some adlayer waters reorient with one OH-group pointing toward the surface (red curve in the inset) and the adlayer structure is partially disordered, with a loss of 0.5 HBs$^{//}$/nm$^2$ (2\% of the total) at -1~V, and of 1.2 HBs$^{//}$/nm$^2$ (6\% of the total) at -2~V. These structural changes are consistent with a previous study employing ab-initio simulations in combination with surface enhanced Raman spectroscopy~\cite{Cheng_NATMAT2019}. By contrast, the water reorientation at positive potentials (see inset) is not accompanied by a disordering within the adlayer, as shown by $\Delta HB^{//}$/nm$^2$, that is larger for potentials of +1~V and +2~V than for 0~V. In agreement with what previously observed at Pt/water interfaces~\cite{Willard_FARADISC2009}, we hence find an asymmetry in the response of the adlayer structure at positive/negative electrodes. However, the changes within the adlayer structure do not significantly alter its hydrophobicity towards the adjacent air/water-like layer, and therefore the cavity formation process. This is measured by the number of inter-layer HBs/nm$^2$ formed between adlayer and air/water-like layer, which remains constant around the value of 4 with a variance of $\pm$0.1 in the whole potential range.\\
\\
\indent It should be noted that in case of metals such as platinum that strongly interact with water, an additional degree of complexity will be provided by the presence of chemisorbed water molecules within the adlayer, as recently shown by Le et al.~\cite{Cheng_SciAdv2020} from ab-initio MD simulations. How much chemisorbed waters can affect the cavity formation mechanism described here, as well as its dependence upon voltage application, is an intriguing question that still remains to be addressed.\\

\subsection{Implications for outer-sphere electrochemical reactions}
We now apply our extension of the LCW theory to a simple model addition reaction:
\begin{equation}
    A + B = C \label{eqn:addition}
\end{equation}
where A and B are small hydrophobic molecules forming cavities of 2.0~\AA~radius, the length-scale of e.g. CO, that react to form a larger molecule of 3.0~\AA~radius, the volume of e.g. ethanol. Addition reactions are key steps for electrochemical routes reducing CO$_2$ to multicarbon products, such as ethanol and ethylene,\cite{HGordon_ACSCAT2018,Fontecave_ACSCAT2020,ChemRev2019_CO2reduction} which are obtained with sufficient faradaic efficiencies on copper electrodes, but not on other metals such as gold~\cite{ChemRev2019_CO2reduction,multicarbon1,multicarbon2}. The hydrophobic hydration contribution ($\delta\mu_{reax}$) to the total free energy of the model addition reaction is given by:
\begin{equation}
\delta\mu_{reax} = \delta\mu_{vC}^{int} - (\delta\mu_{vA}^{int} + \delta\mu_{vB}^{int})   \label{eqn:addition_cavity}
\end{equation}
where $\delta\mu_{vA}^{int}$, $\delta\mu_{vB}^{int}$, $\delta\mu_{vC}^{int}$ are the free energy cost to form the cavities for the two reactants and the product, respectively. The free energy cost to form a 2.0~\AA~radius cavity in the stable outer-shell position is equal to 41~meV (1.6 $k_BT$), while it amounts to 216~meV (8.3 $k_BT$) for the larger 3.0~\AA~radius cavity. Therefore, $\delta\mu_{reax}$ equals to 134~meV. A slightly larger product forming a cavity of 3.5~\AA~radius would result in $\delta\mu_{reax}=$277~meV.\\
\indent The unfavourable hydration of large hydrophobic molecules at the Au/water interface thus imposes a free energy penalty to outer-sphere addition reaction steps, while it can promote elimination reactions (which follow the opposite path: $C = A + B$). Such energetic penalty is not negligible, since $\delta\mu_{reax}$ of our model reaction is of the same order of magnitude as the reaction free energies theoretically determined for common addition reaction steps~\cite{HGordon_ACSCAT2018,Fontecave_ACSCAT2020}. Therefore, our results suggest that hydrophobic hydration could actively contribute to the preferential formation of monocarbon over multicarbon products from CO$_2$ reduction at gold electrodes.
\section*{Discussion and Conclusions}\label{conclusion}
The hydration of small hydrophobic solutes forming cavities of 2.0-3.5~\AA\ radius at an electrified Au(100)/water interface has been investigated from classical MD simulations. In a wide potential range relevant for electrochemical applications, the most stable position for the hydrophobes at the interface is found to correspond to an outer-sphere adsorption, i.e. occupying an inter-layer region separated from the metal by the water adlayer. In contrast, inner-sphere adsorption is hindered by the high free energy cost to form a cavity within the adlayer, where the density of HBs is 5 times higher than in the inter-layer region. Sub-nanometric heterogeneity hence exists for the adsorption of hydrophobic species at the metal/water interface, with an unfavourable spot within the adlayer and a favorable position in the inter-layer region.\\
\indent The interfacial water network is proposed to play a crucial role: while it promotes the accumulation of small hydrophobic solutes of sizes comparable to CO or N$_2$, the free energy cost to hydrate larger hydrophobes, above 2.5 \AA~radius, is greater at the interface than in the bulk. In stark contrast with the Au(100) case, even cavities of 3 \AA~radius were shown to be enriched at Pt/water interfaces, as a result of the stronger chemisorption of water on the platinum surface~\cite{Limmer_PNAS2013,Michaelidis_PRL2003}. Therefore, the more a metal surface interacts strongly with water, inducing a very ordered adlayer structure with a low density of inter-layer HBs, the more hydrophobic hydration is promoted, increasing the limit size for accumulation of solutes at the interface. In the limit scenario where a purely hydrophobic interface is formed between the adlayer and the air/water-like layer, i.e. with zero inter-layer HBs, the enrichment of both small and large solutes will be favoured at the interface with respect to bulk~\cite{Chandler_JPCB2010,Garde_fluctuationsREVIEW2011}. Interesting perspectives are opened by the possibility to trigger the adsorption of hydrophobic species at the metal/water interface as a function of the degree of ordering within the adlayer. Moreover, since density fluctuations in liquid water can deviate significantly from the canonical spherical shapes~\cite{Hassanali_JPCA2017,Patel_JPCB2014}, the sub-nanometric heterogeneity observed in the direction normal to the surface at the gold/water interface suggests that not only the volume but also the shape of the cavity could affect hydration free energies.\\
\indent In summary, when evaluating the free energy cost for hydrophobic hydration in the inter-layer region as a function of the cavity size, the volume-dominated and surface-dominated regimes as described by the Lum-Chandler-Weeks theory~\cite{Chandler_JPCB1999} for the bulk can be identified. At the interface, however, the transition from the 1$^{st}$ to the 2$^{nd}$ regime occurs for hydrophobes of $\sim$ 3 \AA~radius, smaller than in the bulk. For such radius, corresponding to the distance between the adlayer and the air/water-like layer, half of the cavity extends in the inter-layer region, while the other half protrudes into the air/water-like layer, inducing the breaking of some HBs. Here, we propose a framework that allows to extend the LCW theory to metal/water interfaces, by rewriting the free energy cost of hydrophobic hydration at the interface as the free energy cost in the bulk plus a size-dependent correction term. While the values for the correction term are specific for a given interface, the approach can be generalized to any other metal/water interface.\\
\indent Finally, we have shown that outer-sphere addition reaction steps suffer from an energetic penalty imposed by the high cost of large cavities formation, while elimination reactions, where a large molecule is decomposed in smaller cavities, are promoted. Our preliminary results pave the way to a fine-tuning of hydration free energies at the interface as an efficient strategy to manipulate the energetics and mechanisms of electrochemical reactions. 

\section*{Computational methods}\label{methods}
A liquid slab composed by 3481 water molecules between two planar Au(100) surfaces (each electrode made of 5 layers, 162 Au atoms each) was simulated using the MetalWalls code~\cite{mw_joss2020}. Three simulations were performed at fixed applied potentials of 0, 1 and 2~V between the electrodes, respectively. 2D periodic boundary conditions were employed, with no periodicity on the direction normal to the Au surface. Box dimensions along $x$ and $y$ directions are of $L_x=L_y=36.63$~\AA. The SPC/E~\cite{spce_jcp1987} model was chosen for water, while Lennard-Jones parameters introduced by Heinz et al.~\cite{heinz_jpcc2008} were adopted for Au(100). Lorentz-Berthelot mixing rules were used to model the interactions between all atoms and a cut-off of 15~\AA\ was used.  Electrostatic interactions were computed using a 2D Ewald summation method, with a cut-off of 12~\AA\ for the short-range part.
The simulation boxes were equilibrated at constant atmospheric pressure by applying a constant pressure force to the electrodes. The electrodes separation was then fixed to the equilibrium value of 78.6~\AA\ (for which the water density in the middle of box corresponds to the bulk value). A second equilibration step of 5~ns has been performed for all simulations in the NVT ensemble with $T=298$~K. After equilibration, production runs of 80~ns each have been collected with a timestep of 2~fs (NVT, $T= 298$~K) and used for the analysis.  \\
\indent The  water-water HBs were computed using the definition of White and coworkers~\cite{Galli_HB_2000} with 
$O(-H) \cdots O$ distance $\le 3.2$~\AA\ and $O-H \cdots O$ angle in the range $[140-220]^\circ$. A second criterion~\cite{Luzar_JCP1993} has been tested to ensure that the results are not biased by the adopted HB definition. For the OH-orientation analysis (inset of Fig.~\ref{volt1}), an OH-group is considered parallel to the Au surface (H-$//$) if it forms an angle of 90$^{\circ}\pm30^{\circ}$ with respect to the normal $z$-direction (oriented from solid to liquid), H-down if it forms an angle $<$~60$^{\circ}$ and H-up otherwise.\\
\indent The free energy cost of cavity formation as a function of the vertical distance from the Au surface has been calculated by sampling the probability ($P_{v}(0,z)$) to find zero water oxygen centers in a spherical probing volume of the chosen radius:\cite{Limmer_PNAS2013}
\begin{equation}
P_{v}(0,z) = e^{-\beta\Delta\mu_v(z)} \label{eqn:free_energy}
\end{equation}
where $\beta=1/k_BT$, $k_B$ being the Boltzmann constant.
\begin{acknowledgements}
The authors thank C. Stein, M. Head-Gordon, M.-P. Gaigeot and K. Tschulik for fruitful discussions. M. H. and S. P. acknowledge financial support by ERC Advanced Grant 695437 THz-Calorimetry. M. S. and A. S. acknowledge financial support by the European Research Council (ERC) under the European Union's Horizon 2020 research and innovation programme (grant agreement No. 771294).
\end{acknowledgements}

\section*{Data availability statement}
The data that support the findings of this study (input files for the simulations, raw data used for the various figures) are available from the corresponding author upon reasonable request.


\bibliography{references}

\section*{Supporting Information}
\indent The $\delta\mu_v(z)$ profiles in Fig. 1C-right of the main text show for all cavities: (i) one first minimum at z-r $\sim$0 \AA, where the cavity contacts the adlayer in one point; (ii) a maximum at z-r $\sim$1.5 \AA, in between adlayer and air/water-like layer; (iii) a second minimum at z-r $\sim$3.0 \AA, where the cavity contacts the air/water-like layer in one point and extends into the bulk. These three configurations are illustrated in Figure S1.\\
\indent In the main text we have commented on how the free energy to form a cavity in case (i) results from the balance between a favorable and an unfavorable term. The favorable term arises from the fact that part of the cavity extends in the inter-layer region (between adlayer and air/water-like layer, where the density of HBs is extremely low), while the unfavorable one arises from the cavity inducing a distortion of the air/water-like layer. As illustrated in Figure S1, when going from case (i) (z=0 \AA) to case (ii) (z$\sim$1.5 \AA), the cavity is moved farther from the adlayer. A smaller portion of the cavity volume now extends into the inter-layer region while a larger portion extends in the region occupied by the air/water-like layer, inducing a more severe distortion of this layer as compared to case (i). Therefore, the balance between favorable and unfavorable terms is shifted toward the latter and the free energy cost of cavity formation increases from case (i) to case (ii). A maximum is reached in the $\delta\mu_v(z)$ profile.\\
\indent When moving the cavity even farther from the adlayer, case (iii), a second stable spot is observed (z$\sim$3.0 \AA). As illustrated in the Figure, in such a stable spot, the formation of a cavity only requires a little distortion of the air/water-like layer, while most of the cavity volume extends into the bulk water region. Increasing the cavity radius, the effect of the cavity on the air/water-like layer is unchanged since the additional empty volume is formed in the subsequent bulk region. Therefore, the free energy cost to form the cavity increases in the same way as it does in the bulk. This explains why $\delta\mu_v(z)$, that is defined as the difference with respect to the bulk reference, does not show any dependence on the cavity radius at z$\sim$3.0 \AA~(Fig. 1C).\\

\begin{figure}[b]
\begin{center}
\includegraphics[width=0.4\textwidth]{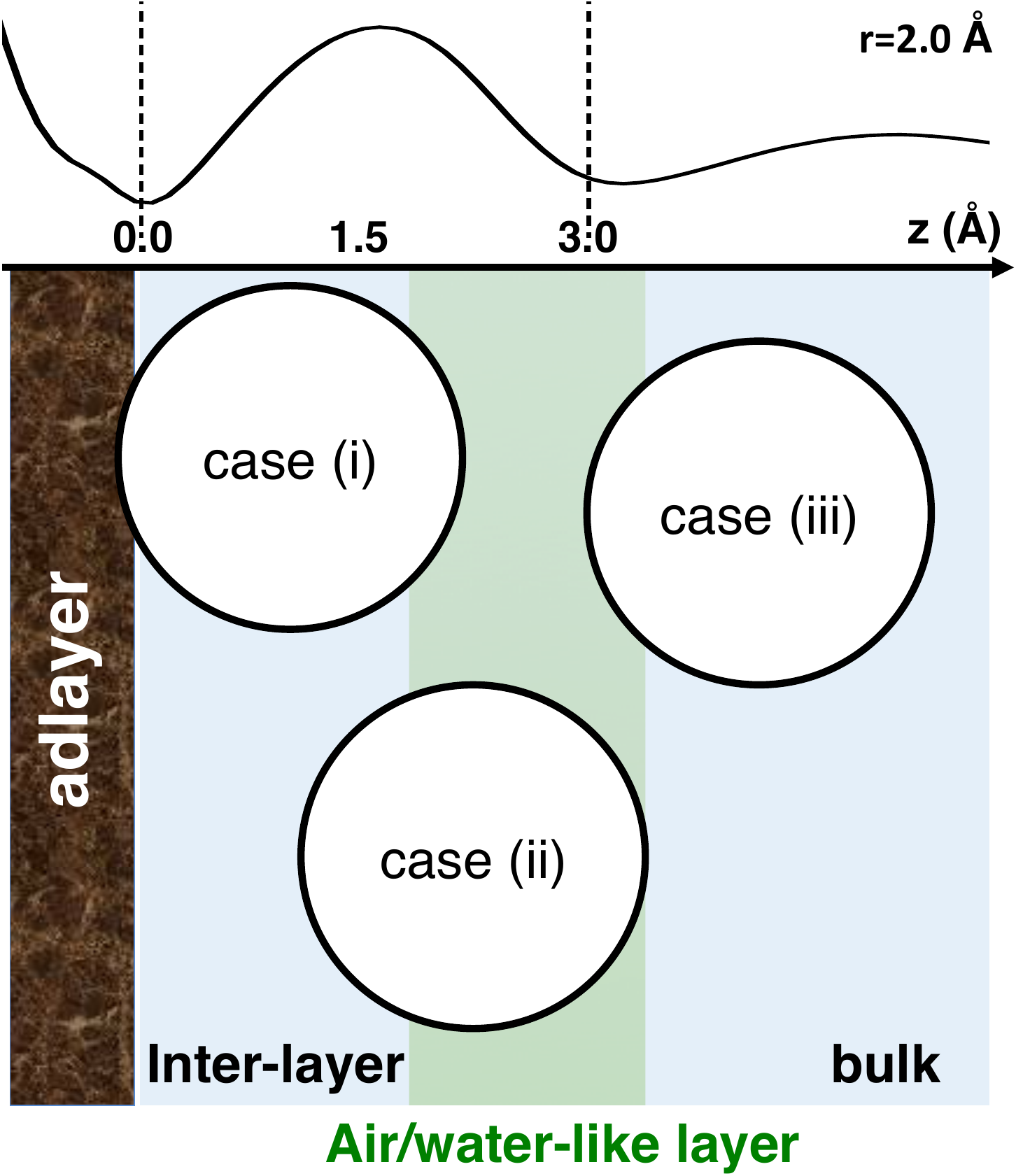}\\
\end{center} 
\figurename{ S1. Schematic illustrating the relevant configurations of ideal hydrophobes (i.e. cavities) hydrated at the Au(100)/water interface, corresponding to the first minimum (i)  (z=0 \AA), first maximum (ii) (z$\sim$1.5 \AA), and second minimum (iii) (z$\sim$3.0 \AA) in the excess solvation free energy $\delta\mu_v(z)$ profiles (reported for a cavity of 2 \AA~radius as an example). $z$ is the distance from the adlayer. The adlayer, inter-layer, air/water-like layer and bulk regions are highlighted.}\label{alinamerda}
\end{figure}
\begin{figure}[t]
\begin{center}
\includegraphics[width=0.47\textwidth]{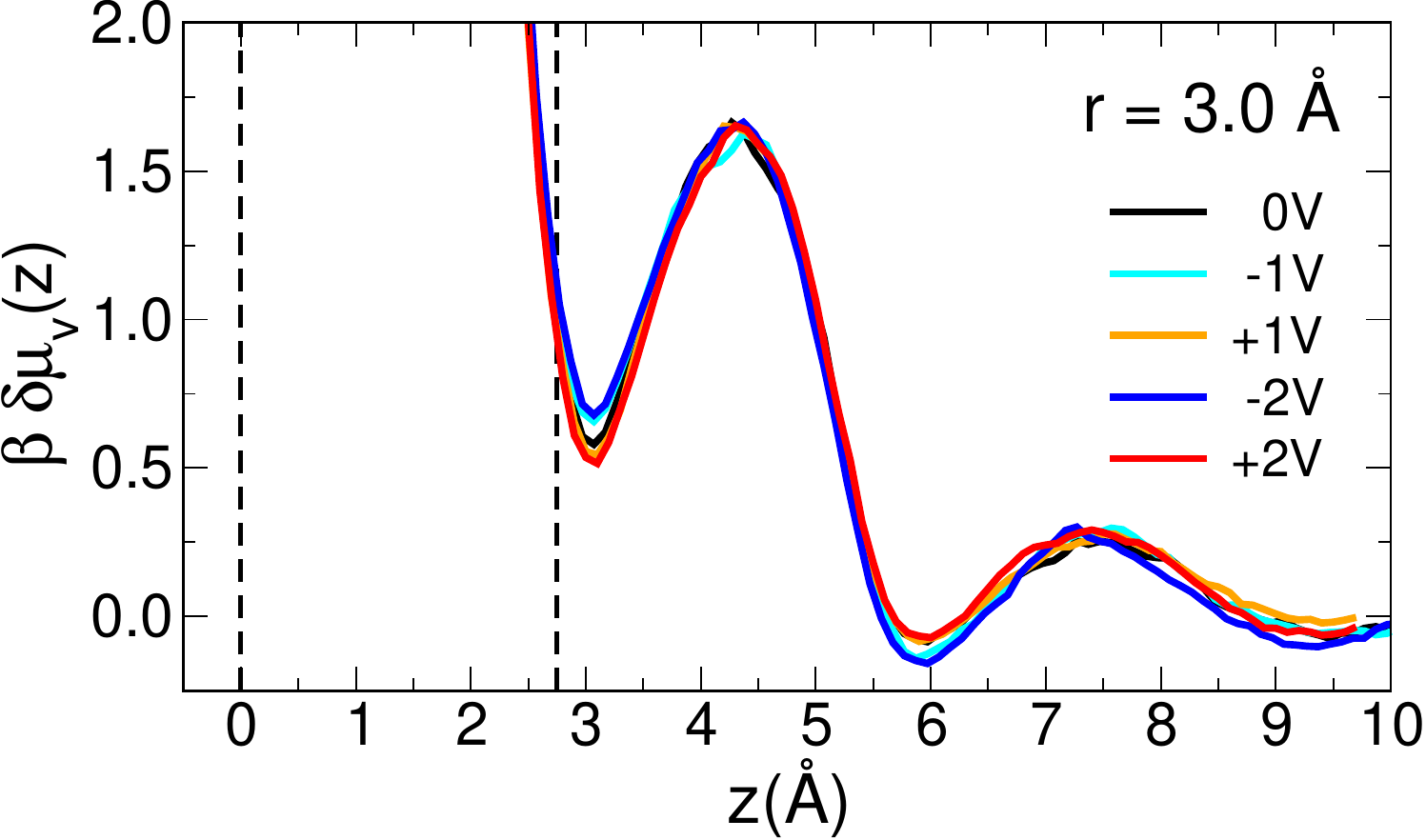}\\
\end{center}
\figurename{ S2. Excess solvation free energy, $\delta\mu_v(z)$, as a function of the applied voltage, for ideal hydrophobes of 3.0 \AA~radius. The "$-$/$+$" notation in the legend refers to the negative/positive electrode. }\label{}
\end{figure}
\begin{figure}[t]
\begin{center}
\includegraphics[width=0.47\textwidth]{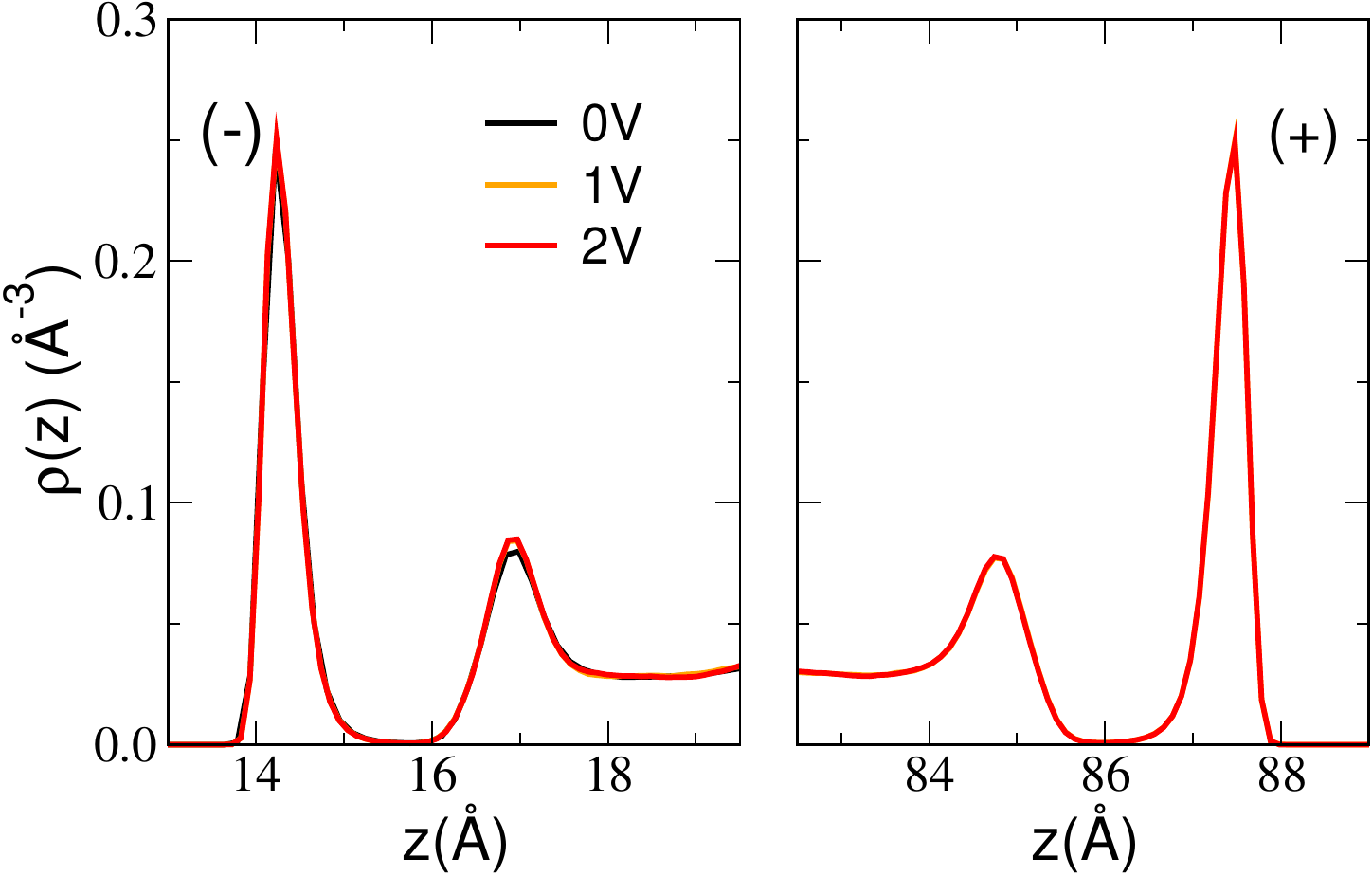}\\
\end{center}
\figurename{ S3. Water density profiles as a function of the vertical coordinate z for various applied voltages. Negative and positive electrodes are shown in left and right panels, respectively.}\label{}
\end{figure}

\end{document}